\documentstyle[epsfig]{article}
\def\b{\bar}
\def\d{\partial}

\def\cD{{\cal D}}

\def\m{\mu}
\def\n{\nu}

\def\~{\widetilde}

\def\bY3{\bar Y_{,3}}
\def\Y3{Y_{,3}}
\def\z{\zeta}
\def\Z{{\b\zeta}}
\def\Y{{\bar Y}}

\def\`{\dot}
\def\be{\begin{equation}}
\def\ee{\end{equation}}
\def\bea{\begin{eqnarray}}
\def\eea{\end{eqnarray}}

\def\fn{\footnote}

\def\mn{{\mu\nu}}

\begin{document}
\title{Kerr-Schild Geometry from Cosmology to Microworld and Space-Time Structure.}
\author{Alexander BURINSKII \\
Theor. Phys. Lab., NSI, Russian Academy of Sciences, \\
Moscow RUSSIA}
\maketitle

\begin{abstract}
The Kerr-Schild (KS) geometry is linked tightly  with
 the auxiliary \emph{flat} Minkowski background. Nevertheless, it describes
 many \emph{curved} space-times and the related physical models, starting from
 cosmology and black holes to the microworld of the spinning elementary particles
 and the pre-quantum structure of vacuum fluctuations. We consider here a KS model
 of the Bubble Universe --  a semi-closed
Universe with a rotating de Sitter (or anti-de Sitter) space
embedded in an external flat space-time. When the solution has two
horizons, it may also be interpreted as an Universe inside a black
hole. In micro-world the KS geometry yields a model of the
spinning particle consistent with gravity and describes a
pre-quantum twistorial structure of space-time with the beam-like
fluctuations of metric consistent with the beamlike fluctuations
of electromagnetic vacuum. These light-like twistor-beams
excitations are consistent with gravity and  generalize the known
pp-wave solutions. Following Wheeler's estimations of the density
of vacuum fluctuations we arrive at the general conclusion that
Universe should be flat and have a zero cosmological constant. It
contradicts predominant doctrine of the Big Bang and expanding
Universe, and enforces us to return to an `effective flat
geometry' filled by the electromagnetic background radiation.
\end{abstract}
\section{Introduction}

General covariance is the main merit of General Relativity and the
main reason of its misinterpretation. The use of different
coordinate systems allows one to treat different embeddings of the
models into the real world which leads to different physical
interpretations of the same solution.

A typical example is the freedom in choice of the radial
coordinate $r$ of the Schwarzschild black hole (BH) solutions, in
which the radial coordinate determines the position of the black
hole horizon, which allows one to use diverse analytic extensions
of this coordinate through the BH horizon. This leads to the
diverse maximal analytic extension (MAE) of the BH geometry and
the appearance of the `black' and `white' holes and many other
topological possibilities which ``... do not have any relation to
the real physics of the black holes.." \cite{MTW}).

The Kerr-Schild (KS) class of metrics has a rigidly fixed
coordinate system related with the flat Minkowski background and
to some extra imprinted causal twistorial structure -- the
congruence of principal null directions. In spite of
 the extreme rigidity and the tight relation to the \emph{flat background},
 the KS metrics include many important curved space-times of General
 Relativity, in particular, they allow one to describe:
 \begin{itemize}
\item rotating black holes and rotating stars (without horizon),
\item de Sitter and Anti de Sitter space-times, and their rotating
analogues, \item a de Sitter (or  AdS) background with an embedded
black hole, \item opposite situations: dS (or AdS) space embedded
inside a black hole, \item charged and rotating black holes and
stars, \item bubble models with a domain wall separating the inner
and external space-times.
\end{itemize}
\noindent The Kerr-Schild metric is given in the Cartesian
coordinates of the Minkowski spacetime ${\rm x}=x^\m=(t,x.y,z) \in
M^4$, and is represented by the simple form $ g_\mn =\eta_\mn + 2H
k_\m k_\n ,$ where $\eta _\mn$ is the auxiliary Minkowski
metric\fn{We use the signature (-+++).}, and the vector field
$k^\m ({\rm x})$ is a null field, $k_\m k^\m =0 ,$  with respect
to the both metrics $g_\mn$ and $\eta_\mn .$

Quantum theory works in flat spacetime and unambiguously suggests
us that the world is flat.
 Principal grounds for this statement follow from the microstructure
of the quantum vacuum -- extremely strong vacuum fluctuations
forming the zero-point field. This huge value of the quantum
vacuum energy is predominate over all other matter and should
determine the structure, and in particular, the curvature of our
Universe. The reality of this field is exhibited by the Lamb shift
and by the
 experimentally confirmed Casimir effect. As it is argued by Wheeler
in \cite{MTW}, the nuclear densities are ~$10 ^{14} g/cm^3 ,$
while estimations of the energy  density for vacuum fluctuations
yield ~ $10 ^{94} g/cm^3 .$  The ratio of these densities may be
compared with the ratio of the energy density of the images on the
cinema screen with respect to the energy density of the screen
itself. The natural assumption is that \emph{the energy of `pictures' should
not curve the screen!}

The KS class of metrics gives us the possibility to combine
consistently the quantum point of view with general relativity:
quantum theory should live on the auxiliary Minkowski background,
while the resulting KS space may be curved in agreement with the
General Relativity.  Twistor structure of the KS geometry,
determined by the discussed bellow Kerr theorem, results in many
other examples of the consistency KS gravity with quantum theory.
In particular, the pp-wave analog of plane waves allows one to
realize in the KS curved spaces the twistorial Fourier transform
\cite{BurPreQ} and creates a pre-quantum KS geometry
\cite{BurExa}. Conformal properties of the KS geometry provide
relations with superstring theory \cite{BurQ,BurGraStr}, which
allows us to build a Kerr-Newman soliton model of spinning
particle consistent with gravity \cite{BurSol}. We note that
twistorial structure of quantum vacuum conflicts with the model of
a closed and expanding Universe. Recall, that the idea of closed
Universe was created by Einstein in 1917, and the principal
motivation for it were the problems in Newton theory --
infiniteness of the gravitational potential for the homogenous
matter density \cite{LL}. At present, this motivation could seem
to be superficial for two reasons: 1) many infinities are known
now for other effectively working and physically accepted
theories, the most important example being QED which is full of
the diverse divergencies, and 2) the real matter distribution in
the Universe is not homogenous, and the potential of the localized
sources should be regularized locally by gravity.

KS metrics allows one to consider the `curved' de Sitter Universe
on the flat Minkowski background with metric $\eta _\mn .$ This
fact was mentioned first by G\"urses and G\"ursay (GG) in
\cite{GG}. In this paper we exploit a generalizations of this
representation.

In sec.2 we consider basic properties and peculiarities of the KS
geometry and also the Kerr Theorem which determines its twistorial
structure. In sec.3. we use the GG form of metric for
regularization of the Kerr-Newman BH solution, and in sec.4 we
consider the based on GG metric model of domain wall bubble, which
corresponds to our interpretation of the semi-closed Universe.

In sec.5 we consider nontrivial consequences of the twistorial KS
structure --  the beamlike exact KS solutions for the
electromagnetic (EM) excitations on the KS background which
indicate a fluctuating twistor-beam structure of the vacuum and
the EM excitations, and finally, in sec.6 we consider the bubble
model of spinning elementary particle based on the regularized
Kerr-Newman (KN) solution.

\section{Structure of the Kerr-Newman solution}
 The KN metric reads
 \be g_\mn=\eta _\mn + 2 H k_\m k_\n , \quad H=\frac {mr -
e^2/2}{\Sigma}, \label{ksm} \ee where \be \Sigma = r^2 +a^2 \cos^2
\theta \label{Sigma}, \ee and $k^\m$ is a null vector field
determined by eq. (\ref{kY}) below. The electromagnetic (EM)
vector potential has the form \be A^\m_{KN} = Re \frac e {r+ia
\cos \theta} k^\m\label{ksGA} ,\ee which is aligned with the
vector field $k^\m ,$ forming the Principal Null Congruence (PNC),
or the Kerr congruence. Gravitational and EM fields are
concentrated near the Kerr singular ring $r=\cos\theta=0,$ where
$r,\theta$ are the oblate spheroidal coordinates (see Fig.1).

\begin{figure}[ht]
\centerline{\epsfig{figure=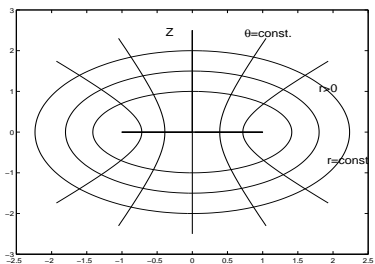,height=3.5cm,width=5cm}}
\caption{The oblate coordinate system $r, \ \theta$.} \end{figure}

In the black hole solutions the Kerr ring is hidden beyond the
horizon. For the large values of angular momentum, in particular,
in the model of spinning particle (sec.6), the Kerr ring is open
and forms a sort of waveguide, or a closed string which generate
zitterbewegung of the Dirac electron. The Kerr ring is a branch
line of the Kerr-Schild geometry into two sheets corresponding to
$r>0$ and $r<0.$

 In the KS representation \cite{DKS}, a few
coordinate systems are used simultaneously. In particular, {\it
the null Cartesian coordinates}
\[ \z = (x+iy)/\sqrt 2 , \
 \Z = (x-iy)/\sqrt 2 , \ u = (z - t)/\sqrt 2 , \
v = (z + t)/\sqrt 2 \] are used for description of the Kerr
congruence in the differential form  \be k_\m dx^\m = P^{-1}( du +
\bar Y d \zeta + Y d \bar\zeta - Y \bar Y dv), \label{kY}\ee via
the complex function $Y(x^\m)=e^{i\phi} \tan \frac \theta 2 ,$
which is a projective angular coordinate on the celestial sphere,
\be Y(x^\m)=e^{i\phi} \tan \frac \theta 2 . \label{Y} \ee

\begin{figure}[ht]
\centerline{\epsfig{figure=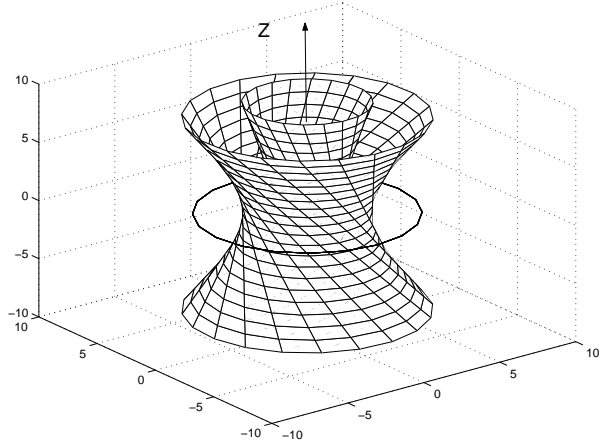,height=6cm,width=8cm}}
\caption{The Kerr singular ring and the Kerr principal null
congruence (PNC).}
\end{figure}

\subsection{Twosheetedness of the KS space-time and
the Kerr theorem} The Kerr congruence (PNC) is controlled by the

 \textbf{Kerr theorem} \cite{DKS,BurNst,KraSte,BurExa,BurPreQ}:
The geodesic and shear-free Principal Null Congruences (PNC) (type
D metrics) are determined by  holomorphic function $Y(x)$ which is
analytic solution of the equation \be F (T^a) = 0 \ , \ee where
$F$ is an arbitrary analytic function of the three projective
twistor coordinates \be T^a =\{ Y,\quad \z - Y v, \quad u + Y \Z
\} .\ee

The Kerr theorem is a practical tool for obtaining  exact
Kerr-Schild solutions, which is performed by the following
sequence of steps:
\[ F (T^a) =0 \Rightarrow F (Y, x^\m) = 0 \Rightarrow Y(x^\m)
\Rightarrow k^\m (x^\m) \Rightarrow g^\mn \] For the Kerr-Newman
solution function $F$ is quadratic in $Y ,$ which yields \emph{two
roots} $Y^\pm(x^\m)$ corresponding to two congruences.

 As a result the two (IN and OUT) congruences determine two sheets of
 the Kerr solution: the ``negative (--)" and
``positive (+)" sheet, where the fields change their directions.

In particular, two different congruences $ k^{\m(+)} \ne
k^{\m(-)}$ determine two different KS metrics $ g_\mn^{(+)} \ne
g_\mn^{(-)} $ on the same Minkowski background. As shown in Fig.2,
the Kerr congruence propagates analytically from the IN to the
OUT- sheet via the disk $r=0 ,$  and therefore, the two KS sheets
are linked analytically, which is conveniently described by the
oblate spheroidal coordinate system $r, \ \theta, \phi ,$ which
asymptotically, for large $r$ goes to the usual spherical
coordinate system. Twosheetedness is the long-term mystery of the
Kerr solution.

The extremely simple form of the Kerr-Schild (KS) metric
(\ref{ksm}) is related with complicate form of the Kerr
congruence, which represents a type of deformed (twisted)
hedgehog. In the rotating BH solutions the usual pointlike
singularity inside the BH turns into a \emph{ closed singular
ring,} which is interpreted as a closed string in the
corresponding models of the spinning elementary particles
\cite{BurSol,BurSen,BurGraStr,BurQ}.

The Kerr singular ring represents some important extra argument
against the singular initiate state of the Big Bang model, since
the pointlike singularity is defocused in the Kerr solution by the
rotation. The singular initiate state should disappear for the
rotating Big Bang model. \fn{An additional argument is that there
is a quantum limit for the superdense state due to the
\textit{vacuum volume Casimir effect} (superdense matter goes into
a pseudovacuum state.) \cite{BurCasim89}.}

The Kerr-Schild geometry gives a new look at the BH horizon. The
sensitivity of the horizon to electromagnetic fields is well known
\cite{BurA}. In particular, the horizons disappear for $|e|>m,$
which is the point of bifurcation. By  $|e|\ge m ,$ the KN
solution is not changed, while all Carter's diagrams (MAE)
disappear. The Kerr coordinates are rigidly linked with a
background Minkowski spacetime, and therefore, the coordinate
system of the KS geometry is decoupled from the position of the
horizon. It allows one to analyze influence of the electromagnetic
(EM) field on the horizon and its deformations by the EM field
\cite{BurA}. The basic KS solutions turn out to be independent of
the position of the horizon and even of its existence.

Similar arguments where given by Misner-Thorne-Wheeler (Sec. 33.2
of \cite{MTW} -- the complicate spacetime of the ``maximal
analytic extension" (MAE) of the Kerr-Newman (KN) black hole
spacetimes does not have any relation to the question on black
holes for two reasons:

a) the most part of the internal KN spacetime takes the
\emph{already collapsed} star, and

b) even the external KN geometry does not give an adequate
description of the real geometry due to the \emph{nonstationary
processes} around BHs.

\section{Phase transition as a regulator of the black hole singularity.}

 Smooth and regular rotating sources of the Kerr-Schild class were
 obtained in \cite{BurBag,BEHM}. They are based on a generalization of the
 KN geometry suggested by the G\"urses and G\"ursey (GG) in \cite{GG}.
 Starting from the KS form of metric
\[g_\mn = \eta_\mn + 2 H
k_\m k_\n ,\] GG  used the generalized form of the function $H ,$
\be H=f(r)/\Sigma, \quad \Sigma=(r^2 + a^2 \cos ^2\theta) ,
\label{HGG}\ee which allows one to suppress the Kerr singular ring
 ($r=\cos \theta =0$) by a special choice of the function $f(r).$

It is assumed that by this deformation, the Kerr congruence,
determined by the vector field $k^\m(x^\m)$ remains lightlike,
$k_\m k^\m =0 ,$ and retains the usual KS form (\ref{kY}).

The GG-form of the metric
 describes the Kerr-Newman BH solution, and also the rotating de
Sitter and Anti de Sitter solutions. Besides, it allows one to
match smoothly the rotating metrics of different types!

The regularized solutions have tree regions:

\textbf{i) the Kerr-Newman exterior}, $r>r_0 $, where $f(r)=mr -e^2/2,$

\textbf{ii) the source region,} $r<r_0-\delta $, where $f(r) =f_{int}$
suppresses the KN singularity, providing smoothness of
the metric up to second derivatives.

\textbf{iii) intermediate region} (domain wall) providing
 a smooth phase transition interpolating between the regions i) and ii).

\noindent To remove the Kerr-Newman singularity, one has to set for the
internal region
\[f_{int}=\alpha r^4 .\]
In this case, the Kerr singularity is replaced by a regular
rotating source, forming internal space-time with a constant curvature, $ R=-24
\alpha ,$ \cite{BurBag}.

 The functions
\begin{equation}
D= - \frac{f^{\prime\prime}} {\Sigma}, \label{Dt}
\end{equation}
\begin{equation}
G= \frac{f'r-f}{\Sigma^2} \label{Gt}.
\end{equation}
determine stress-energy tensor  in the orthonormal tetrad
$\{u,l,m,n\}$  connected with the Boyer-Lindquist coordinates,
\begin{equation}
T_{ik} = (8\pi)^{-1} [(D+2G) g_{ik} - (D+4G) (l_i l_k -  u_i
u_k)]. \label{Tt}
\end{equation}
In the above formula, $u^i$ is a timelike vector field given by
$$
u^i=\frac 1{\sqrt{\Delta\Sigma}}(r^2+a^2,0,0,a) ,
$$
where $\Delta= r^2 +a^2 - 2f  ,$ \cite{BurBag,BEHM}.

 This expression shows that the matter inside the source is
separated into ellipsoidal layers corresponding to constant values
of the coordinate $r$, and each layer rotates with angular
velocity $\omega(r)= \frac {u^{\phi}}{u^0}=a/(a^2+r^2)$. This
rotation becomes rigid only in the thin shell approximation,
$r=r_0$. The linear velocity of the matter w.r.t. the auxiliary
Minkowski space is $v=\frac {a \sin \theta}{\sqrt {a^2 + r^2}}$,
so that on the equatorial plane $\theta =\pi /2$, for small values
of $r$ ($r\ll a $), one has $v \approx 1$, that corresponds to an
oblate, relativistically rotating disk.

The energy density $\rho$ of the material satisfies
$T^i_ku^k=-\rho u^i$ and is, therefore,  given by
\begin{equation}
\rho = \frac{1}{8\pi} 2G. \label{rhot}
\end{equation}
Two distinct spacelike eigenvalues, corresponding to the radial
and tangential pressures of the non rotating case are
\begin{equation}
p_{rad} = -\frac{1}{8\pi} 2G=-\rho, \label{pradt}
\end{equation}
\begin{equation}
p_{tan} = \frac{1}{8\pi}(D+ 2G)=\rho +\frac{D}{8\pi}. \label{prtt}
\end{equation}
In the exterior region function $f$ must coincide with Kerr-Newman
solution, $f_{KN} = mr -e^2/2$.

There appears a "domain wall" -- a transition region
 placed in between the boundary of the matter object (the de Sitter interior) and the external KN gravi-electro-vacuum solution.
  This transition region is described by a smooth function $f(r)$ interpolating
between the functions $f_0(r)$ and $f_{KN}(r)$.

\section{Domain wall bubble with two horizons as a regularized black
hole and as a cosmological model}

The resulting source may be considered as a bubble or bag filled
by a special matter with positive ($\alpha>0$) (or negative
$\alpha < 0$) energy density. This bubble source may be naked or
covered by the horizon. If such a source is covered by the
horizon, it will be seen by external observer as an usual black
hole, since the external observer cannot check  presence of the
source behind the horizon. On the other hand this black hole is
regularized, since its singular source is replaced by the domain
wall bubble.\fn{The shell bubble source of the KN solution with
flat interior was first suggested by L\'opez \cite{Lop}.}

{\bf Graphical analysis for the case $\alpha
>0$}
\begin{figure}[ht]
\centerline{\epsfig{figure=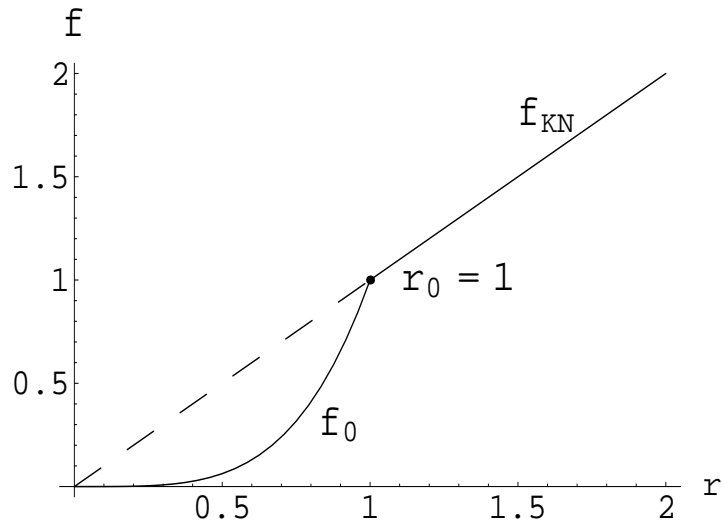,height=7cm,width=10cm}}
\caption{\small Position of phase transition $ r_0 $ as an
intersection of plots $ f_0 (r) $ and $ f_{KN} (r) $. Uncharged
source, $ \alpha > 0 $, arbitrary units.}
\end{figure}

The case $\alpha >0$ corresponds to de Sitter interior and
uncharged source.
 There is only one intersection between $f_0(r)=\alpha r^4$ and
$f_{KN}(r)=mr$. The position of the transition layer will be $r_0
=(m/\alpha)^{-1/3}$. The second derivative of the corresponding
interpolating function will be negative at this point, yielding an
extra contribution to the positive tangential pressure in the
transition region.

{\bf Positions of the horizons}

 The external region is described by the KN
electro-vacuum solution, and the transfer from the external KN
solution to the internal region (source) may be considered as a
phase transition from `true' to `false' vacuum. The point of phase
transition $r_0$ is determined by the equation
$f_{int}(r_0)=f_{KN}(r_0).$
 The point of intersection $r_0$ corresponds to balance of the mass.

\begin{figure}[ht]
\centerline{\epsfig{figure=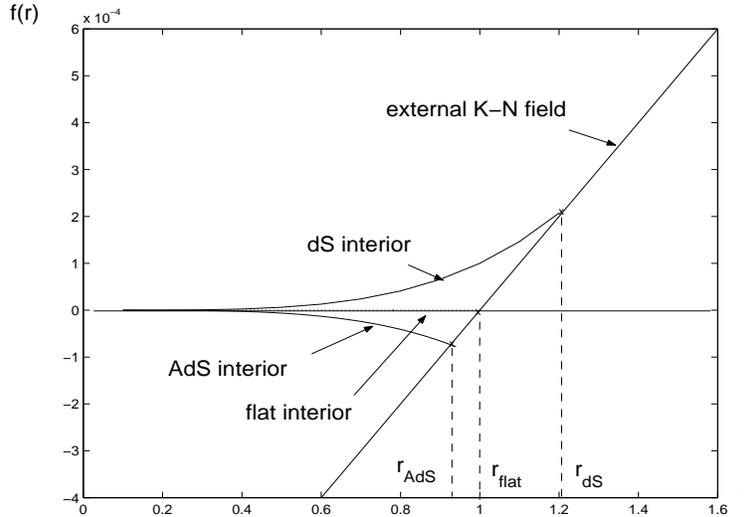,height=7cm,width=10cm}}
\caption{The phase transfer from the external KN solution $f(r)=f_{KN}$ to the
internal solution $f(r)=f_{int}$ for $r<r_0$. }\end{figure}

Consider first a {\bf non-rotating case, $\Sigma=r^2$:}
$\alpha=8\pi \Lambda/6 .$ The inner  space-time has a constant
curvature $R=-24 \alpha.$ There is a de Sitter interior for
$\alpha >0,$ and anti de Sitter interior for $\alpha <0$. The
interior is flat if $\alpha =0. $ The energy density of the source
is $ \rho = \frac 1 {4\pi} (f'r -f)/\Sigma^2 , $ the tangential
and radial pressures are $ p_{rad}=-\rho, \quad p_{tan}=\rho -
\frac 1 {8\pi}f''/\Sigma .$

{\bf Transfer to rotating case.} One has to set $ \Sigma=r^2 +a^2
\cos^2 \theta , $ and consider $r$ and $\theta$ as the oblate
spheroidal coordinates.

The Kerr source represents a disk with the boundary $r=r_0$ which
rotates rigidly. In the coordinate system corotating with the
disk, the matter of the disk looks  homogenously distributed;
however, because of the relativistic effects the energy-momentum
tensor increases strongly in equatorial plane near the boundary of
the disk.

{\bf The positions of the horizons} for the usual charged and
rotating black hole solution are determined by the relation $
\quad r_\pm = m \pm \sqrt{m^2 -e^2 -a^2} $.
 The positions of the horizons for the considered bubble-source
 are determined by graphical analysis represented in Fig.5.
  One sees that similar to the case of the charged BH, there appear two
  horizons around the position of the domain wall $r_0$: the external horizon
  $r_+$ is just similar to the usual BH horizon and hides the domain wall from
  external observer. At the same time, the internal horizon is similar to
  cosmological horizon and hides the domain wall from
the observer living inside the domain wall. It gives the model of
an Universe inside a BH.

\begin{figure}[ht]
\centerline{\epsfig{figure=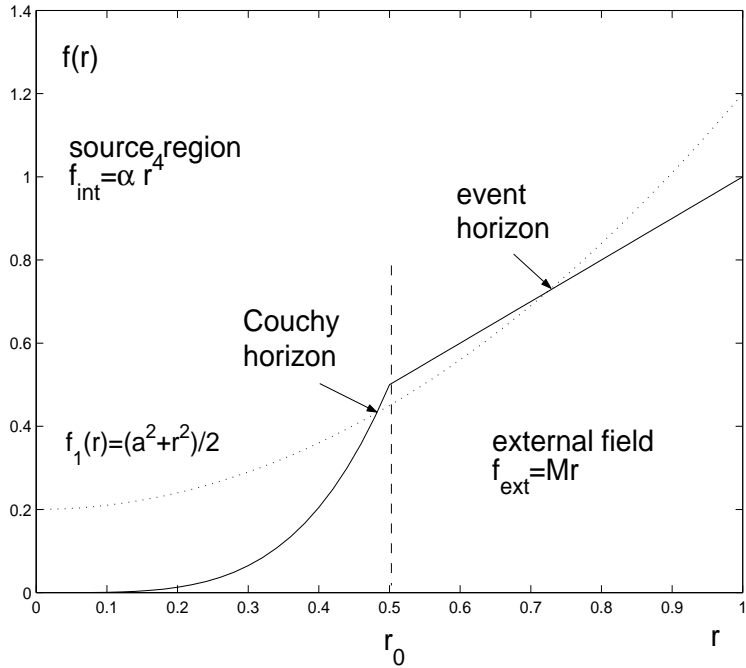,height=9cm,width=10cm}}
\caption{``De Sitter'' source of the external Kerr-Newman field.
The dotted line $f(r)=(r^2 +a^2)/2$ corresponds to the allowed
positions of the horizons.}
\end{figure}

If the rotation increases, the horizons disappear and the domain
wall bubble is exhibited as a rotating stars or another object
consistent with general relativity. There appears principal new
feature that the KN singularity is regularized and the naked
rotating solution does not break the principle of cosmic
censorship. In particular, the model of such a source was
considered in \cite{BurSol,BurBag,Lop,Isr,Bur0,DirKN,TN,Dym,David}
by the treatment of the Kerr-Newman (KN) solution as a  model of
spinning particle consistent with gravity. It was shown that  the
system of chiral fields (involving the Higgs field) is able to
form a positive potential interpolating the corresponding phase
transition from internal to external supersymmetric states.
Therefore, the bubble is formed by a domain wall separating the
external and internal regions of the bubble.

We are going to the principal point of our interpretation of the
regularized black hole solutions with two horizons $r_+$ and
$r_-$. The observer, who is positioned outside the external
horizon $r_+ ,$ may consider this horizon as the usual  horizon
related with this black hole, which hides the interior of the
black hole from him.  At the same time, the observer, who is
positioned \emph{inside the internal horizon} of the regularized
black hole solution $r_-$, sees the usual almost flat world, and
in fact he lives in a de Sitter space for which \emph{the horizon
$r_-$ is the `cosmological' horizon.} He will never  know what
happens behind this horizon, and what is the structure of the full
solution beyond the horizon. Therefore we arrive at the model of
the {\it Universe inside the BH solution,
 or inside of an elementary particle}. Ideas of this type on the junction of the de Sitter-type interior through the transition layer to the Schwarzschild geometry, or
 about the  geometry of Universe isomorphic to a semi-closed world, were considered
 earlier by Klein (1961), Zeldovich-Novikov (see \cite{ZeldNov} and  refs. therein)
 and were expressed the most close to our point of view by Frolov, Markov and
 Mukhanov in \cite{FMM}. However, note that the considered above KS representation
 results in the model which is  rotating and connected with the flat
 background, and also, it describes the smooth phase transition without an appeal
 to quantum arguments.

\section{Fluctuating twistor structure of the KS geometry.}

Analysis of the exact electromagnetic solutions on the KS
background confirms that gravity conflicts with the usual plane
waves. It has been shown
 that there are no
 smooth harmonic solutions on the KS background, and elementary excitations of the black-holes and their consistent
back-reaction on the KS metric form twistor-beams -- singular
beams supported by twistor null lines of the Kerr congruence,
\cite{BurExa,BurPreQ,BurA}. The twistor-beams are similar to laser
beams, and turn asymptotically into singular pp-waves -- a type of
fundamental strings --  gravitational analog of the fundamental
heterotic strings of the string field theory \cite{BurGraStr}.
 The beam-like basic excitations of the BH have important consequences for the
 physics of the BHs.  The BH horizons are extra sensitive to the electromagnetic (EM)
 field \cite{BurA}\fn{In particular, it is known that position of the BH horizons depends on the value of the BH charge and the horizons disappear at all for the strongly charged BH solutions.}. The electromagnetic beams have very strong back reaction to metric and deform topologically the horizon. It has been shown \cite{BurA,BurExa} that
 the EM excitations of the BH solutions
 create the EM beams which penetrate the horizon, creating
 there the holes which connect the BH interior with external region, \cite{BEHM2}.

Since the horizon is extra sensitive to electromagnetic
excitations, it should also be sensitive to the vacuum
electromagnetic excitations (which are exhibited classically as a
Casimir effect), and therefore, the created by vacuum excitations
twistor-beam pulses shall also perforate horizon, producing a
fine-grained structure of fluctuating microholes which allow
radiation to escape fr4om the interior of the black-hole,  Fig.3.
It yields a semiclassical mechanism of the BH evaporation.

\begin{figure}[ht]
\epsfig{figure=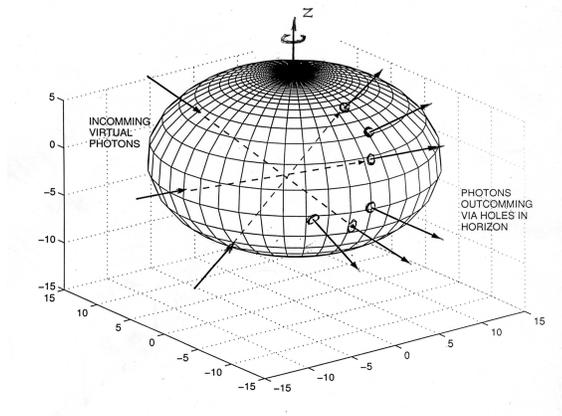,height=6.5cm,width=7.5cm} \vspace*{-7mm}
 \caption{Excitations of a black hole by a
 weak electromagnetic field create twistor-beams perforating the BH horizon
 by fluctuating micro-holes.}
\end{figure}

Another important effect related with the twistor-beams is
exhibited for the multiparticle Kerr-Schild solutions,
\cite{wonder,Multiks}. In agreement with the Kerr theorem, the
generating function $F(Y,x^\m)$ of the Kerr congruence takes for
the $n-$particle Kerr-Schild solutions the form of the product of
partial functions $F_i$ for the $i$-th particle, \be F(Y,x^\m)=
\prod^n_i F_i(Y,x^\m). \ee Since each $F_i$ is quadratic in $Y,$
the resulting, solution of the basic equation \be F(Y)=0 ,
\label{F0} \ee  has $2n$ roots $Y^{\pm}_i(x^\m), \ i=1,2,..n ,$
 and the KS geometry turns out to be \emph{multivalued and multisheeted} twistorial space-time,
 having independent congruences of the twistor null lines on different sheets.
 In general, these congruences ``do not interact and do not feel each other".
However, if $Y^{\pm}_i(x^\m)=Y^{\pm}_k(x^\m)$ there appear the
multiple roots in the basic equation  (\ref{F0}),
 which create extra null singular lines which join the $i$-th and $k$-th particles.
 For each two particles $i$ and $k ,$ there are exactly two such twistor null
 lines  - one is directed from $i$-th to $k$-th  ($Y^+_i(x^\m)=Y^-_k(x^\m)$)
 and another one from $k$-th to $i$-th ($Y^+_k(x^\m)=Y^-_i(x^\m) .$)
\begin{figure}[ht]
\centerline{\epsfig{figure=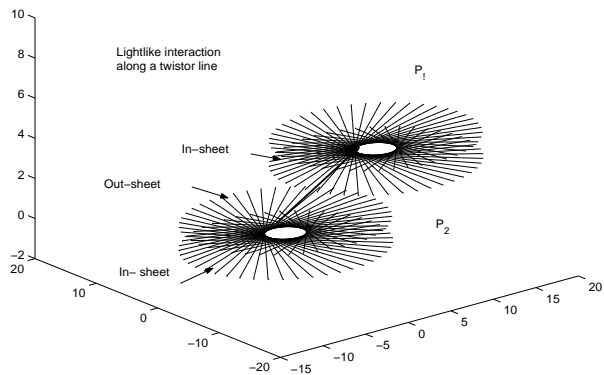,height=6cm,width=8cm}}
\caption{Four sheets   of the Kerr space formed by the null
congruences of two particles. The sheets are analytically
matched  via a common twistor null line.}
\end{figure}
  As a consequence, each two particles in the Universe turn out to be connected by two
singular lines (pp-wave strings), supported by twistor lines which
are common for these particles. The electromagnetic excitations
consistent with gravity  form pp-waves propagating along these
 twistor lines, realizing a type of ``photon exchange". The complex analyticity of these
 pp-waves breaks down at the points of their emanation (which corresponds to
their creation) and at the points of their termination
(corresponding to their annihilation).\fn{The creations and
annihilations occur at the matter sources, like the discussed
above domain wall bubbles.}
 The pp-waves have a very strong back reaction
 to the metric and the space-time which turns out to be covered by fluctuating
 twistor-beams (twistorial alternative to graviton). Therefore, the pp-wave twistor-beams turn out to
 be the basic elementary excitations of the KS space-time, representing an analog of the
 electromagnetic vacuum fluctuations in configuration space consistent with gravity.
 The twistorial structure of the KS geometry forms a multisheeted  background of the fluctuating twistor-beams,
 creating a fluctuating pre-quantum KS geometry \cite{BurPreQ,BurA}.
 Due to the conformal structure of the KS geometry, there appears a way to derive
 procedure of  quantization  from the formalism of the operator product expansion (OPE).

\section{Regularized KS geometry as a background of the  electron}

  Quantum theory states that electron is pointlike and
structureless. For example, Frank Wilczek writes: {\it
``...There's no evidence that electrons have internal structure
(and a lot of evidence against it)"}.  Leonard Susskind writes
similarly: the electron radius is  {\it``...most probably not much
bigger and not much smaller than the Planck length.."}, i.e.
($l_p= m_p G/c^2=1.6\cdot 10^{-33} cm $).

However, the observable parameters of the electron, mass $m$,
charge $e,$ spin $s$ and the gyromagnetic ratio $g=2$ (\cite{Car})
indicate unambiguously that its gravitational background should be
the KN solution. Because of the very large spin of the electron,
$a=J/m >> m ,$ the BH horizons disappear, so, the KN background
with the electron parameters is not a black hole! It is not flat
and has the above discussed non-trivial twosheeted topology
created by the Kerr singular ring.
 For parameters of the electron the KN metric is almost flat
everywhere and gravitational field is concentrated near the Kerr
singular ring. This ring can be identified with a closed gravitational
string \cite{BurGraStr,BurStr,IvBur}
 which is very similar to the Sen solution for the closed heterotic string.\fn{The
field around the Kerr-Sen solution to low energy string theory  \cite{Sen} is similar to the Sen solution for fundamental heterotic string \cite{BurSen}.}
Elementary excitations of the Kerr string form the lightlike pp-wave traveling waves \cite{BurGraStr,BurQ,BurA,Bur0} accompanied by axial twistor-beams  along twistor lines of the Kerr congruence.
 The Kerr closed string has the Compton radius, and traveling waves
along the string create the known effect of zitterbewegung giving rise to the Dirac
theory of electron \cite{BurGraStr}.

The singular stringlike source of the KN metric can be regularized
by the considered above mechanism of a phase transition, which
turns the KN solution into a gravitating soliton model, forming a
rotating bubble source covering the former KN singular ring
\cite{BurSol,BurGraStr}. \fn{The bubble source turns out to be
highly oblate because of the relativistic rotation.} The interior
of the bubble is filled with the Higgs field which regularizes the
EM field, expelling it from the bubble. The twisted EM field forms
a closed Wilson (Bohm-Aharonov) loop along the sharp boundary of
the oblate bubble, which interacts with the phase of Higgs field,
producing the quantized angular momentum \cite{BurSol,BurGraStr}.
It is remarkable that quantization of the angular momentum appears
as a consequence of the semi-classical Higgs mechanism of the
broken symmetry.

The pointlike experimental exhibition of the electron may be
explained as a result of its relativistic rotation \cite{BurQ}.
The lightlike twistor null lines of the Kerr congruence are
tangent to the Kerr ring forming a heterotic string with the
light-like circular current. This string is relativistically
sliding along itself and may look as a pointlike particle because
of the \emph{relativistic contraction}. One can show  that
scattering of this string by \emph{the real} photons  should
exhibit a pointlike structure because the real  photons (with the
lightlike interval $ds^2=0$) will have only one point of the
incidence with the lightlike string. First of all we note, that
the lightlike interval, $s_{12}^2=0 ,$ between a position of the
EM source $x_1$ and the point of its incidence with the  Kerr
string $x_2$ exists only in equatorial plane of the Kerr
spacetime\fn{It is well known from the analysis of the photon
trajectories in Kerr geometry.} Then, one sees that there exists
only one incident point in the equatorial plane. This point is the
tangent point to the lightlike direction of the Kerr string.

However, it was argued in \cite{BurQ,BurGraStr} that the Compton
size of this string may apparently be observed in the experiments
with the \emph{deeply virtual} Compton scattering, which paid
great attention recently as a novel regime of the "nonforward"
Compton scattering \cite{Rad,Ji}. On the other hand, as it was
recently noticed in \cite{BurGra},  the relativistic invariance of
Kerr's angular momentum, $J=ma ,$ should result in the abrupt
shrinking of the Compton size $a$ for relativistic electrons.

\section{Conclusion}
The based on twistors KS geometry displays wonderful universality
penetrating all the regions of theoretical physics from cosmology
and black holes to the structure of elementary particles and
twistorial microstructure of spacetime. Twistorial structure of
the KS geometry creates new solutions and new effects which remove
contradictions between gravity and quantum systems, representing a
new perspective to the problems of quantum gravity.
\section*{Acknowledgements}
 I am very thankful to Prof. Ruggero Santilli for invitation to attend
 this remarkable
 conference and for financial support. I am also thankful to Prof.
 Theo Nieuwenhuizen for the thorough reading the manuscript and
 worth remarks, and also
 to Prof. Dirk Bouwmeester for invitation to Leiden University
 where this paper was finished.

\section*{Appendix A. Supersymmetry and cosmological constant}

 Supersymmetry appears in many different contexts in theoretical
physics. The recent experiments at LHC rejected supersymmetry in
its primitive interpretation. In our opinion, the idea that each
particle has its \textbf{\emph{superpartner}} is incorrect, since
it is based on the wrong idea that in a theory with unbroken
supersymmetry, for every type of the boson there exists a
corresponding  fermion with the same mass and internal quantum
numbers, and vice-versa. In fact, theory of supersymmetry
\cite{WessBag} states it only on the field level for a
supermultiplet of the fields. Supersymmetry states only that there
should be equal number of the bosonic and fermionic degrees of
freedom. On the same grounds we believe that the Higgs boson will
never be obtained, since the Higgs field is only one of the fields
of supermultiplet - it is not particle, although some kind of the
special type of soliton formed from the Higgs field may exist and
may be obtained in future. In this respect I consulted  Prof.
Santilli, and his opinion is that inconsistency of supersymmetry
(conflict with unitarity) may be retained on the level of field
theory too.\fn{The Santilli theory generalizes algebraic grounds
of supersymmetry including generalization  to irreversible
processes.}

Meanwhile the theory of superfields is based on the Hermitian
Lagrangians and is extremely elegant,  possessing exclusive
effectiveness for description of the models with broken symmetry
including the supergravity models.

 Relationship of the \emph{zero cosmological constant} with supersymmetry  was demonstrated
 by Zumino in \cite{Zum}. The zero point energy is formed
 of the mutually compensating bosonic and fermion vacuum modes.
 The contribution is
\[ N\frac 1{2(2\pi)^3}
\int \sqrt {\vec p ^2 + m^2} d^3 p ,\]  \noindent with N the
number of one-particle states of excitations. It holds that $N>0$
for tensor fields and $N<0$ for spinor fields.

\noindent Each particle is formed by a supermultiplet of the
fields, and vacuum energies compensate among the various physical
\emph{fields of a supermultiplet -- not among the particles!}

\noindent In particular, electron is not a pure fermion. It
contains the bosonic EM field  and the mass creating gravitational
fields. In the corresponding field theory it is formed as a
soliton built of a supermultiplet of the fields including the
electromagnetic field, spinor field, gravity and a system of
scalar fields, see sec.6.

\section*{Appendix B. Field model of phase transition}
An elegant analytic description of the phase transition can be
obtained in the frame of Domain Wall model based on the
supersymmetric system of the chiral fields. The considered in
\cite{BurSol} model represents a generalization of the Abelian
Higgs model used by Abrikosov, Nielsen and Olesen (ANO)
 for description of the vortex string in superconducting
media \cite{NO}. The ANO Lagrangian is \be {\cal L}_{NO}= -\frac
14 F_\mn F^\mn + \frac 12 (\cD_\m \Phi)(\cD^\m \Phi)^* + V(r).
\label{LNO}\ee For the field model coupled with gravity, the
derivatives $ \cD_\m = \nabla_\m +ie A_\m $ have to be considered
as covariant ones. In \cite{BurSol} the system of chiral scalar
fields  $\Phi^{(i)} = \{\Phi, Z, \Sigma \}, i=1,2,3$ is
considered, and the Lagrangian takes the form \be {\cal
L}_{matter}= -\frac 14 F_\mn F^\mn -\frac 14 B_\mn B^\mn + \frac
12 \sum_i(\cD^{(i)}_\m \Phi^{(i)})(\cD^{(i) \m} \Phi^{(i)})^* + V
\label{L3} ,\ee where $F_\mn = A_{\m,\n} - A_{\n,\m} , \quad B_\mn
= B_{\m,\n} - B_{\n,\m} .$ The covariant derivatives are \be
\cD^{(1)}_\m =\nabla_\m +ie A_\m , \ \cD^{(3)}_\m =\cD^{(3)}_\m
=\nabla_\m + ie B_\m , \ \cD^{(2)}_\m = \nabla_\m . \label{covD}
\ee

 Adapting the scalar multiplet $\Phi^{(i)} = \{\Phi, Z, \Sigma \}$ to
 the bubble source, we consider the first field $\Phi^1 (x) $ as the usual
 Higgs field, which should vanish outside the bubble and acquire a nonzero
 vacuum expectation value (vev) inside the bubble, giving a mass to the
 electromagnetic field $A^\m ,$ which enforce  to vanish the EM field
 inside the bubble. The scalar field $\Phi^3 (x) $ is considered as another
 (dual) Higgs field which has opposite behavior, taking
zero vev inside the bubble and a nonzero vev outside. The related
with $\Phi^3 (x) $ gauge field $B^\m$ should acquire a mass
outside the bubble, which keeps it confined inside the bubble. In
the bubble-soliton model \cite{BurSol} the field $A^\m$ forms a
closed string concentrating near the sharp boarder of the bubble
in the equatorial plane. By introduction of the second gauge field
$B_\m ,$ this model approaches the Weinberg-Salam theory, and
there is expected the appearance of the second `electroweak'
closed string related with the field $B^\m.$ The scalar field
$\Phi^2 (x) = Z $ is assumed to be uncharged and used for
synchronization of the phase transition. This very special
mechanism of the phase transition should be arranged by a special
behavior of the potential $V(x),$ which is provided by the
algorithm known from theory of supersymmetry \cite{WessBag}.

\noindent {\bf Supersymmetric phase transition.} In accord with
theory of supersymmetry \cite{WessBag},  the potential $V$ which
determines the character of phase transition and the vacuum states
are described analytically from a \emph{superpotential $W$,} which
is analytical function of the chiral fields, $\Phi^i(x).$ The
 potential is determined by the relations \be V(x) = \sum _i |\d_i W|^2 .\label{VW}
\ee
 Vacuum states, which provide minimum of the potential $ V,$ are determined
 by the condition $\d_i W
=0 .$

\emph{The form of the superpotential used in \cite{BurSol} } is
\be W= Z(\Sigma \bar \Sigma -\eta^2) + (cZ+ \m) \Phi \bar \Phi
.\label{W}\ee where $c, \ \m, \ \eta $ are real constants. It
generates a domain wall with the described above necessary
properties. The supersymmetric potential given by  (\ref{VW}) and
(\ref{W}) is non-negative and interpolates between the external
\emph{vacuum} state $V_{(ext)}=0$ and internal \emph{pseudovacuum}
state $V_{(int)}=0 .$ {\emph{There appear two supersymmetric
vacuum states separated by a domain wall}:
 the vacuum state inside the bubble, $V_{(int)} =0$ for $r<r_0 ,$
with the vev solutions $ \Phi ^i(x)=\Phi ^i_{(int)} ,$ and
 the external vacuum state, $V_{(ext)} =0$ \ for $r>r_0$ with the
vev solutions $ \Phi ^i=\Phi ^i_{ext} .$

\end{document}